\documentclass[twocolumn,reprint,preprintnumbers,amsmath,amssymb,prl]{revtex4}

\usepackage{graphicx}% Include figure files
\usepackage{dcolumn}% Align table columns on decimal point
\usepackage{bm}% bold math
\usepackage{color}
\usepackage{hyperref}
\hypersetup{colorlinks=true,citecolor=red}

\hypersetup{colorlinks=true,citecolor=red}

\begin{document}
	\title{InSe as a case between 3D and 2D layered crystals for excitons}
	%\title{InSe between 3D and 2D layered crystals }
	\author{W. Desrat$^1$}
	\author{T.V. Shubina$^2$}
	\author{M. Moret$^1$}
	\author{A. Tiberj$^1$}
	\author{O. Briot$^1$}
	\author{B. Gil$^{1,2}$}
	\affiliation{$^1$ Laboratoire Charles Coulomb (L2C), Universit\'e de Montpellier, CNRS, Montpellier, FR-34095, France}
	\affiliation{$^2$ Ioffe Institute, 26 Politekhnicheskaya, St Petersburg 194021, Russia}
	\date{\today}

	\begin{abstract}
		We demonstrate the successive appearance of the exciton, biexciton, and P band of the exciton-exciton scattering with increasing excitation power in the photoluminescence of indium selenide layered crystals. The strict energy and momentum conservation rules of the P band are used to reexamine the exciton binding energy. The new value $ \geq20 $ meV is markedly higher than the currently accepted 14 meV, being however well consistent with the robustness of excitons up to room temperature. A peak controlled by the Sommerfeld factor is found near the bandgap ($\sim1.36$ eV), which puts the question on the pure three-dimensional character of the exciton in InSe, which has been assumed up to now. Our findings are of paramount importance for the successful application of InSe in nanophotonics.
	\end{abstract}

	%\pacs{78.20.-e, 78.45.+h, 78.55.-m, 71.35.-y}
	\maketitle
	
%%%
	
	Many interesting optical phenomena in 2D crystals are associated with the free excitons (X), trions, and biexcitons or excitonic molecules (M) created due to the strong Coulomb interaction between particles. For instance, the hybrid configuration of biexcitons, which can involve states from different valleys and dark ones, was discovered in monolayers of transition metal dichalcogenides \cite{Sie2015,Hao2015,Nagler2018}. Further, the intervalley scattering can produce the dark biexciton states which can be radiative and whose lifetime is comparable with that of bright excitons \cite{Danovich2017}. 
	
	Similar investigations into monochalcogenides are at the very beginning. 
	The most studied are GaSe and InSe - strongly anisotropic layered crystals consisting of tetralayers of Se-Ga(In)-Ga(In)-Se bound by weak van der Waals forces. They have indirect and direct bandgaps respectively in bulk, and the type of band structure changes to the opposite in the limit of a single tetralayer (see \cite{Rybkovskiy2014} and references therein). In InSe, such crossover with the formation of a "Mexican hat" valence band is extended to a dozen of tetralayers\cite{Rybkovskiy2014,Mudd2016}. This stimulated us to study excitons in multilayered InSe. 
	
	The stacking sequence in InSe leads to different polytypes. In the $\gamma$-polytype ($C_{3v}$ point group) and $\beta$-polytype ($D_{6h}$ point group), the direct transitions between the uppermost valence and the lowermost conduction bands are fully allowed only for light polarization $E\parallel\textit{c}$ axis \cite{Kuroda1980,Segura2018}. The optical transition with $E\perp\textit{c}$ seems possible through a weak spin-orbit interaction by taking into account the spin of the states\cite{Kuroda1980,Segura2003,Errandonea2005}.  
	
	\begin{figure*}[t]
		\includegraphics[width=0.8\textwidth]{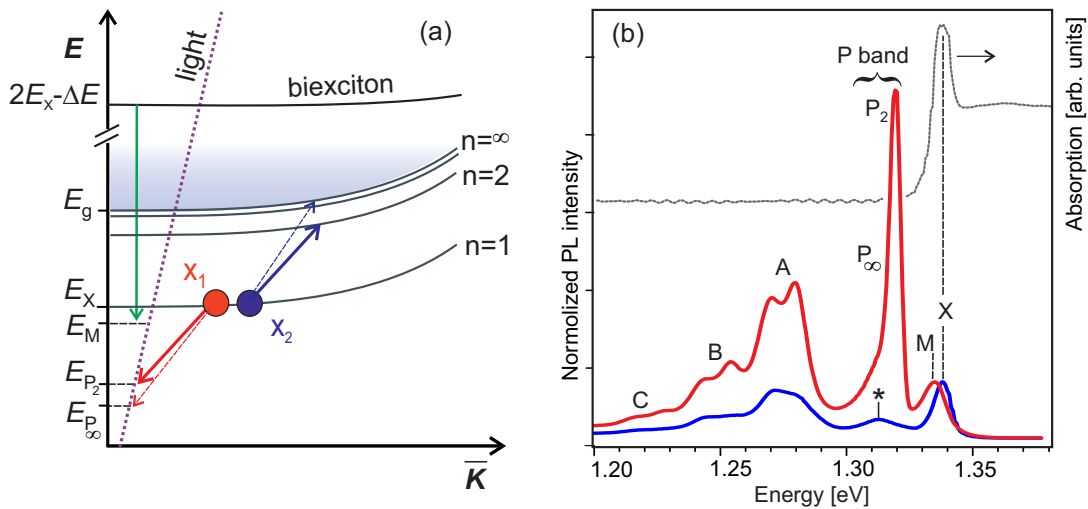}
		\caption{\label{fig1}  \textbf{Exciton, biexciton and exciton-exciton scattering}. (a) Schematic representation of the biexciton decay and the X-X scattering which involves the $n=2$ and $n=\infty$ states shown by solid and dashed arrows respectively. The exciton level notation is given for the 3D case; it should start from $n=0$ for the 2D case. The characteristic energies are marked on the vertical axis ($\Delta E \approx E_X-E_M$).
		(b) Photoluminescence spectra (linear scale) measured under low ($50$~W/cm$^2$, blue line) and high ($0.3$~MW/cm$^2$, red line) excitation powers in undoped InSe.  An absorption spectrum (grey line) is shown for comparison. }
		%\label{twocolumnfigure}
	\end{figure*}	
	
	In 1968, Andriyashik \textit{et al.}\cite{Andriyashik1968} detected two peaks in the absorption spectrum of bulk InSe, which they ascribed to the ground and first excited exciton states. They derived the direct energy gap $E_g\simeq1.36$~eV and the exciton binding energy $R_X\sim37$~meV. The latter is almost twice higher than in GaSe, where the perfect series of exciton peaks was recorded up to $n=3$ \cite{LeToullec1980}. 
	In 1978, Camassel \textit{et al.} \cite{Camassel1978} have revised the InSe data using similar transmission measurements. The energies of the observed peaks led to $R_X=14.5$ meV  and $E_g$=1.353 eV. Even smaller values were obtained by differential magneto-optical measurements\cite{Merle1978}. 
	The modelings in all these experimental data were done using the three-dimensional (3D) theory of allowed direct excitonic transitions. The seeming applicability of this model was a crucial argument in favor of the 3D character of excitons in InSe, instead of the expected two-dimensional (2D) type for layered crystals.
	
	Here, it is worth mentioning about some factors which can influence absorption spectra, such as the Sommerfeld factor which provides a peak at an absorption edge due to the effects of excitons within the continuum\cite{Belenkii1983}, and the interference peaks which can appear in the region of relatively low absorption above the ground exciton energy. Nevertheless, the 1978' values are frequently used up to now and the 3D exciton case is commonly accepted.
	
	As an alternative way for the determination of these fundamental parameters, we are considering the non-linear photoluminescence (PL) processes. In general, with increasing excitation power one should successively observe the appearance of the PL lines of the exciton (X), biexciton (M), and the so-called P band of exciton-exciton (X-X) scattering. In the latter, one of the exciton is scattered into a photon, while the other is scattered into an excited state or to the continuum \cite{Klingshirn1975}. Such series have been observed in many semiconductors, however, up to now this was never reported for the monochalcogenides. Importantly, the energies of emitting photons are strictly dependent on the exciton binding energy as shown in the scheme in Fig.~\ref{fig1}a, which depicts both P band constituents - $P_2$ and $P_{\infty}$. 
	
	The X transitions  exhibit usually a linear dependence of PL intensity on excitation power. 
	In contrast to that, both biexciton recombination cascade and  X-X scattering are characterised by a quadratic power dependence. To confirm the biexciton formation a superlinear dependence of the PL intensity on power is enough. For the inelastic scattering between two excitons, the exact quadratic dependence is typical because this process dominates over others at high excitation power. At the end this X-X scattering results in the emergence of stimulated emission (SE)\cite{Klingshirn2007}. Formally, the SE may be asserted when and only when the PL intensity shows a super-quadratic dependence on power.
	
	A feature related to the biexciton was observed in GaSe, about 2 meV below the X line\cite{Dey2015}, by means of the nonlinear two-dimensional Fourier transform technique. A similar study of InSe did not report on such an observation \cite{Dey2014}. Both biexcitonic and P bands were revealed by power-dependent PL studies in layered lead iodide crystals \cite{Ando2012}, whereas only P and SE bands were reported for bulk GaSe \cite{Mercier1975, Cingolani1987} and InSe \cite{Cingolani1982}. The super-quadratic dependence expected for SE was published for GaSe only, where it likely involves the direct states situated only $25$~meV above the indirect ones. The band structure of InSe with a valence band splitting of hundreds meV is hardly consistent with such a scenario. However, it makes this compound suitable for the investigations of excitons and exciton complexes, as well as the exciton character (3D or 2D) missed in the previous studies.
	
	Here, we  demonstrate the perfect  series of X, M, and P emission bands in InSe which obey the theoretically predicted laws. We use the strict energy and momentum conservation rules of the P band constituents to determine the exciton binding energy and estimate the bandgap width. The comparison of the obtained parameters with the currently accepted ones and the consideration of the temperature robustness of excitons as well as the appearance of the Sommerfeld peak in absorption cast serious doubts on the purely 3D character of the exciton in InSe.
	
%%%
\bigskip
\textbf{\large{Results}}
\smallskip
\\
\textbf{Photoluminescence}.
The experiments were carried out on samples freshly cleaved from bulk InSe grown by the Bridgman-Stockbarger process. We investigated two kinds of samples:  without intentional doping (naturally n-type), called further as "undoped",  and p-type Zn-doped InSe. Both samples possess good structural quality (see the Supplementary Material~\cite{SM}). 
	
Figure~\ref{fig1}b presents the PL spectra of undoped InSe measured at $T=10$~K with low and high excitation powers. For the sake of demonstration, these spectra are normalised to the maximal intensities of the peaks marked as X(M). The peaks A-C have been assigned to different defect-related transitions (see the Supplementary Material~\cite{SM}).  The peak marked by a star (*) is likely related to a donor-like defect. Here we focus on the shorter-wavelength lines with the maxima at $1.338$~eV (X), $1.335$~eV (M), and 1.320 eV (P-band). The line at $1.338$~eV perfectly corresponds to the free exciton recombination~\cite{Camassel1978,Merle1978,Abha1982}. It is also well matching the ground exciton peak detected in the absorption spectrum of a thin sample. Its well-defined energy ($E_X$) is the starting point for further analysis.  
	
\begin{figure}[t]
		\includegraphics[width=1\columnwidth]{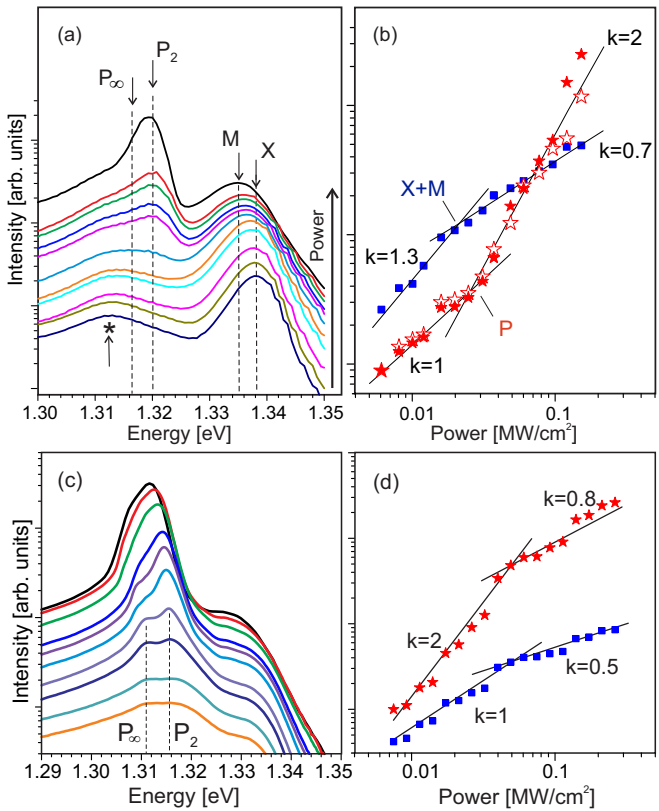}
		\caption{\label{fig2} \textbf{Excitation power dependence of PL}. (a,c) Selected PL spectra measured at different excitation powers (log scale) in undoped (a) and doped (c) InSe samples. (b,d) Power dependencies of integral PL intensity shown for the joint X+M peak (blue) and P band (red) in the undoped (b) and  doped (d) InSe. Black lines show fittings with the exponents marked nearby. In (b), solid and open red stars present the P-band data obtained, respectively, as for a whole band and as a sum of two components.  }
	\end{figure}

In our samples, the M and X peaks cannot be separately resolved because of their closeness and broadening. However, we notice that the joint X+M peak shifts by $\sim$3 meV to the lower energy with increasing the excitation power (Fig.~\ref{fig2}a). The band renormalization cannot explain this behavior since the energy of the near P band is very constant. The scattering to higher-energy states should be very sensitive to this process. This joint peak exhibits a superlinear dependence on pump power with an exponent $k=1.3$ (Fig.~\ref{fig2}b). We assume that it is the result of the admixture of a biexciton (although an exciton complex or trion cannot be completely excluded now). The superlinear growth occurs up to the threshold of intense X-X scattering. Beyond that the exponent decreases down to $k=0.7$. Note that the competition between these two processes -- biexciton formation and X-X scattering -- tends always towards the latter \cite{Nakayama2010}. 
	
The attribution of the $1.320$ eV band to the X-X scattering P-band is proved by its power dependence shown in Fig.~\ref{fig2}b. The P band comprises two overlapping components, P$_2$ and P$_\infty$. They can be well separated due to a significant energy distance between the first excited state and a free-particle bandgap corresponding to the maximum of the P$_\infty$ line (see Fig.~\ref{fig1}a). In the PL spectrum measured at $T=10$~K the basic peak of the P band is situated $18$~meV below X. 
The principal question is which of the components, P$_2$ or P$_\infty$, produces the maximum of the P band. 
For GaSe, the dominant peak was ascribed to $P_2$ ~\cite{Mercier1975}. We also incline to this option based on its lineshape. At low temperature, the P$_2$ should display the Lorentzian lineshape, because the scattering has the well defined level $n=2$ as a final state (in 3D notation). On the contrary, the P$_\infty$ line should be widened due to the large number of possible final states. The higher-energy part of the P-band in Fig.~\ref{fig1}b can be perfectly fitted by a Lorentzian, while its lower energy part deviates markedly from that. 
	
The determination of the P$_\infty$ energy in undoped InSe is complicated by the donor-related line (*) situated nearby, which come close to P$_\infty$. To get rid of this problem we have investigated a Zn-doped InSe sample, where the possible donors are compensated by the p-type impurity. Spectra measured in this sample contain the peaks of both components (Fig.~\ref{fig2}c) separated by $\sim$5 meV. These peaks exist up to the onset of the exciton-electron scattering, well defined by means of a P-band shift and the deviation of the power dependence from the quadratic law (Fig.~\ref{fig2}d). 
	
\begin{figure}[b]
		\includegraphics[width=0.85\columnwidth]{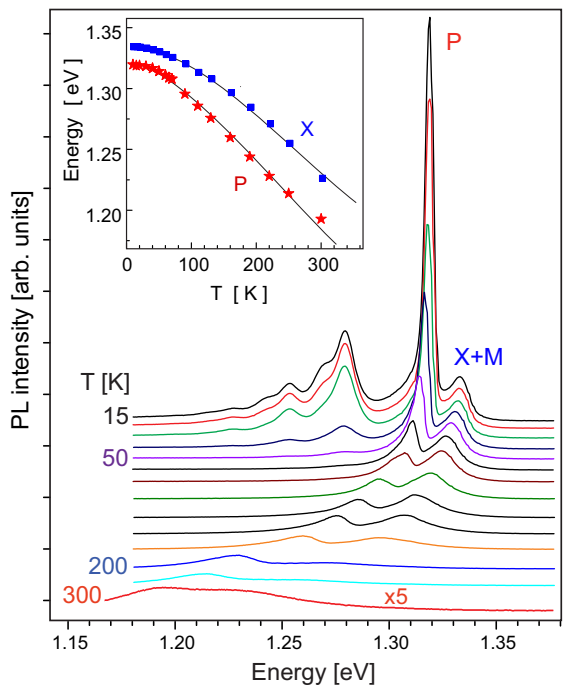}
		\caption{\label{fig3}  \textbf{Temperature dependence of PL}. PL spectra measured at high excitation power in the undoped InSe sample at different temperatures (linear scale). The inset presents the dependencies of the X and P energies vs temperature.}
	\end{figure}

Figure~\ref{fig3} show the PL spectra measured under high pumping power in a wide temperature range. Whereas the defect-related lines quench fast at the temperature of $\sim$60~K, both the exciton-related emission and P band survive almost up to room temperature. This would correspond to an exciton binding energy of about $20$~meV. These emission bands are significantly broadened at temperatures above $80$~K which indicates the involvement of phonons to the recombination process (see the Supplementary Material~\cite{SM} for details). The increase of the gap between the X and P lines (inset of Fig.~\ref{fig3}) is related to the gradually increasing contributions of the exciton-electron scattering and electron-hole plasma recombination to the formation of the P band\cite{Klingshirn2007}. 
	
	\begin{figure*}[t]
		\includegraphics*[width=1\textwidth]{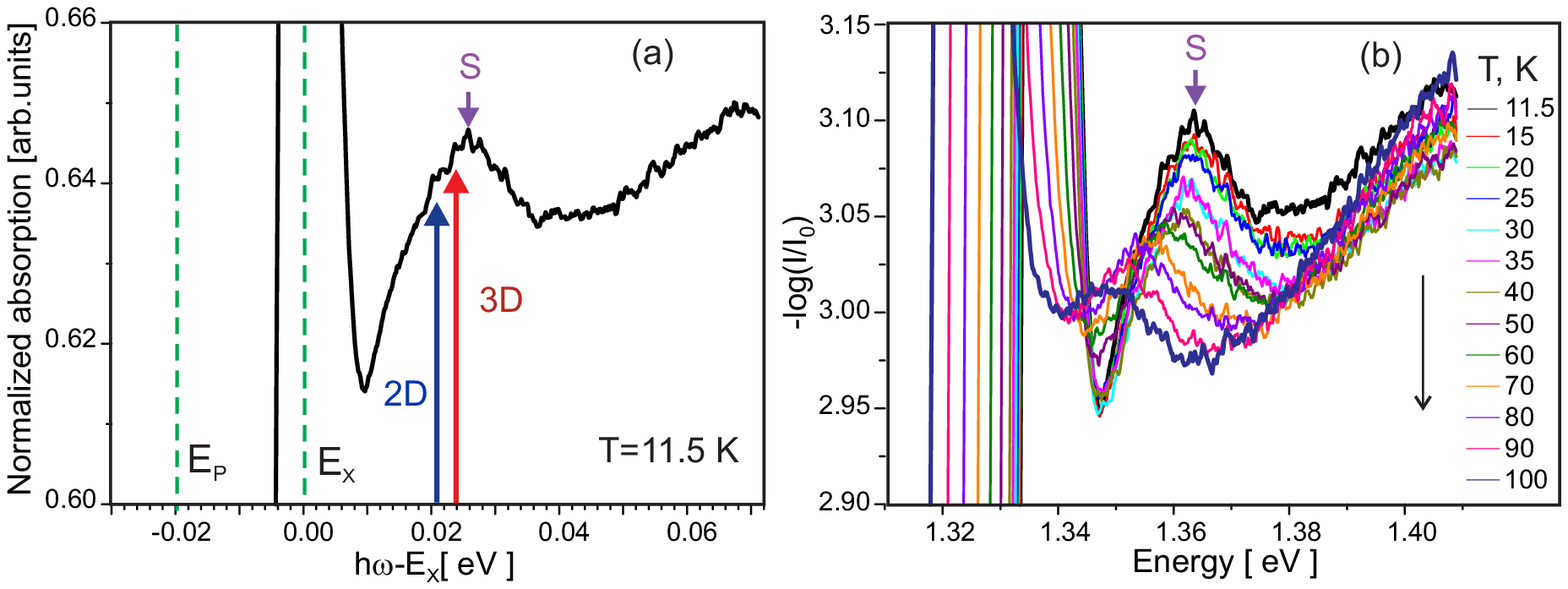}	
		\caption{\label{fig4} \textbf{Sommerfeld peak}. (a) An absorption spectrum measured at $11.5$~K normalized to the intensity of the X peak, whose energy is taken as zero. The vertical axis scale is chosen to focus on the Sommerfeld peak (S). The bandgap energies derived from the P-band analysis are shown for the 3D (red arrow) and 2D allowed (blue arrow) cases. (b) The absorption coefficient measured at different temperatures.}
		%\label{twocolumnfigure}
	\end{figure*}

\textbf{Absorption}.
The absorption spectra measured in a thin ($40$~$\mu$m) undoped InSe sample exhibit also the X peak surviving up to high temperatures ($\geq250$~K). Details on the absorption spectra fitting using the Urbach rule are given in the Supplementary Material~\cite{SM}. Despite our best efforts in measuring samples of different thicknesses, we have failed to find the definite peaks of excited exciton levels as reported in Ref.~\cite{Camassel1978}, which is the almost unique description in the literature. Instead of that, a smooth peak ("hump") was recorded $\sim25$ meV above the X peak. The intensity of this peak is much lower than that of X but it is well visible in an enlarged scale (see Fig.~\ref{fig4}a). Its shift and quenching with temperature follow those of the ground exciton peak (Fig.\ref{fig4}b). 

%%%
\bigskip
\textbf{\large{Discussion}}
\smallskip
\\
\textbf{Rydberg energy}.
The radiative decay of free excitons takes place only when their momentum is located inside a rather narrow light cone which results in the existence of a "bottleneck". This leads to the collection of excitons, which stimulates the formation of exciton complexes and the enhanced exciton-exciton scattering governed by strong energy and momentum conservation rules.
The following consideration relies upon the energy distance between the excitonic levels, which is affected by the 3D or 2D character of the exciton. 
	
For the 3D case, the energies of the P band constituents can be expressed via the exciton binding energy $R_X$ as~\cite{Klingshirn1975}
	\begin{equation}\label{eq1}
	\hbar\omega^{3D}=E_g(T)-R_X\left(2-\frac{1}{n^2}\right)-3\delta k_{B}T, ~n=1,2,3...
	\end{equation}
At low power and temperature when the last term  is negligible, the  $P_2$ energy  counted from the energy $E_{X}$ is 
	\begin{equation}\label{eq2}
	(E_X-\hbar\omega)^{3D}=\frac{3}{4}R_X.
	\end{equation}
For $E_X-\hbar\omega=18$ meV derived from experiments, the bandgap energy is $24$~meV above $E_X$, i.e. at $1.362$~eV. The possible correction by the term $3\delta k_{B}T$ is insignificant because the PL maximum does not markedly shift either with increasing power, as would be expected in the case of additional laser heating, or with temperature in the range $10\div20$~K. Among different points on the sample surface, the energies of the PL lines vary within $\pm0.5$~meV~\cite{SM}. Thus the estimated error in the $R_X$ determination is about $\pm1$meV.
	
In the Zn-doped InSe sample we have measured an energy difference $\hbar\omega(P_{2})-\hbar\omega(P_{\infty})\sim5$ meV which does not depend on temperature at all. For the 3D case, it  leads to $R_X \simeq20$~meV which may be considered as a lower limit of the exciton Rydberg energy in InSe. Note that $E_g\sim1.36$~eV is in agreement with the published experimental data dispersed in a wide range \cite{Gurbulak1999, Duman2007} and with our experimental absorption data as well. We also underline the closeness of the Rydberg values obtained in different InSe samples.

Let us now consider the ideal 2D case. For allowed transitions, the energies of bound excitons are $E_g-R_X/(n+1/2)^2$, with $n=0,\,1,\,2,\dots$.  The scattering involving the $n = 0$ and $n = 1$  levels will produce a photon with energy 
	\begin{equation}\label{eq3}
	(E_X-\hbar\omega)^{2D}_{A}=\frac{32}{9}R_X.
	\end{equation}
The bandgap will be situated at the energy $4R_X$, i.e. $\sim20$ meV above $E_X$. Forbidden transitions in the 2D case start from $n=1$ only~\cite{Ralph1965,Shinada1966}. The respective bandgap would be $4R_X/9=28$~meV above $E_X$. We discard this possibility because the transitions in InSe are at least partly allowed. 
	
\textbf{Sommerfeld peak}.
We have marked in Fig.~\ref{fig4} the values of $E_g$ predicted for the 3D and 2D ideal cases.   
They are close to each other and both are situated below the maximum of the discovered peak. Note that the Coulomb interaction not only ensures the formation of a series of discrete levels below the ionization edge, but also changes the wave functions of the continuum states above this edge~\cite{Shinada1966}. It leads to an absorption peak controlled by the Sommerfeld factor which writes $C(j)=2/(j+1/2)^3$ with $j=0\dots\infty$. The contribution of the $j^\text{th}$ continuum excitonic state (unbound exciton) to the absorption drops exponentially with its increasing detuning from the edge. As a result the Sommerfeld peak appears very close to the bandgap energy. The most pronounced peak occurs in 2D confinement where it can locally increase the step-like 2D absorption by a factor of 2. 
To form such a peak the exciton states must be allowed~\cite{Ralph1965,Belenkii1983} which is fulfilled in InSe. Thus, the clear diagnostic of the Sommerfeld peak is one of the ways to determine $E_g$. In addition, it might indicate the 2D character of the excitons. 
	
\textbf{InSe exciton between the 3D and 2D cases}.
The biexciton binding energy $R_{M}$, determined as the difference $\Delta E \approx E_X-E_M$, is about $0.1R_X$ for bulk and $\sim0.2R_X$ for 2D structures~\cite{Singh1996}. In InSe where $\Delta E \approx 3$ meV, the ratio $R_{M}/R=0.15$ is consistent with the intermediate case between 2D and 3D. 
The recent calculations of the band structures in layered monochalcogenides have demonstrated that the specific ring-shaped valence band is saved up to 4-6 tetralayers in GaSe and GaS, while this critical thickness in InSe is extremely high, approaching 28 tetralayers \cite{Rybkovskiy2014}. In relation to that it is worth noting that monochalcogenide crystals contain a lot of stacking faults. The stacking disorder confines excitons within a finite number of tetralayers \cite{Forney1977} that can promote their 2D character.  

On the other hand, the value of the exciton binding energy $R_X\approx20$ meV obtained in this work, is noticeably higher than the previously reported one ($14$ meV~\cite{Camassel1978, Merle1978}). It is comparable with that in GaSe~\cite{LeToullec1980}, where the exciton was considered as 3D because of the out-of-plane expansion of the wave functions due to the contribution of Se orbitals~\cite{Belenkii1986}. In InSe the higher anisotropy of all parameters can emphasize this effect~\cite{Gomesdacosta1993}. However, the 3D concept does not explain why the excited states are not observed in the absorption spectra, while in the 2D case they are simply too close to the ionisation edge and cannot be spectrally resolved.

Recently, electron energy loss spectroscopy  (EELS) has shown that the exciton peak at $1.3$~eV in InSe does not exhibit any dispersion, i.e. the exciton band structure is flat apart from the zone center \cite{Politano2017}. The small exciton dispersion in InSe due to its high ionicity was predicted long ago~\cite{Depeursinge1978}. In such dispersionless conditions, the effective in plane masses are heavier\cite{Segura2018} and the exciton Rydberg $R_X\propto\mu/m_0$ should be higher. 
This, along with strong anisotropy in $k$~\cite{Ralph1965}, does not give a chance for the exciton states in InSe to match either a perfect $1/n^2$ 3D series, or the one described as $1/(n+1/2)^2$ for extreme two-dimensionality. Additional theoretical studies are clearly needed to elucidate the excitonics in InSe.

\textbf{Conclusion}.
We have detected the successive appearance of the free exciton, biexciton, and P band in InSe using PL measurements with increasing excitation power up to $\sim0.3$~MW/cm$^2$. The well-defined energies of the P band constituents with respect to the X peak allow us to estimate the values of the exciton binding energy and bandgap energy. It is worth noting that our data are obtained using rather transparent modeling which can be easily checked. We assume that the difference between the previously published data and ours occurs not only due to the limited adaptation of ideal models developed for 3D and 2D excitons, but also due to the existing uncertainty in real band structure and exciton wavefunctions in some layered crystals, such the InSe case considered here. Our findings including the enhanced stability of the exciton in the high temperature range and the first observation of the exciton complex in InSe are of paramount importance for future applications of this compound in nanophotonics and quantum optics. 

\bigskip
\textbf{\large{Methods}}
\smallskip
\\
\textbf{Optical measurements}. The cleaved samples (millimeter-sized parallelepipeds) were mounted on the cold finger of a closed cycle helium cryostat with the possibility of temperature variation in the $10-350$~K range. A continuous $15$~mW red laser ($650$~nm) was used to measure low-intensity PL. A Q-switched green laser ($532$~nm) was used for high power pumping. The pump intensity was varied using neutral filters. The PL was detected by a Hamamatsu InGaAs photomultiplier cooled at $77$~K. Besides PL, we measured transmission in a thinner ($\sim40$~$\mu$m) InSe sample with a halogen lamp from $T=10$~K up to $350$~K.\\
\\
\textbf{Structural characterization}. The structural quality of the samples was characterized by x-ray diffraction and Raman studies. Both methods confirmed the good quality of the samples. The experimental details are given in the Supplementary Information.

\bigskip
\textbf{\large{References}}
\\

\bigskip
\textbf{\large{Acknowledgements}}
\smallskip
\\
T.V.S. and B.G. acknowledge the partial support of the Government of the Russian Federation (Project No. 14.W03.31.0011 at the Ioffe Institute). We thank M.A. Semina and M.M. Glazov for very fruitful discussions.

\bigskip
\textbf{\large{Author contributions}}
\smallskip
\\
The optical experiments were carried out by W.D., M.M., A.T. and O.B. The x-ray diffraction was performed by M.M. The theory was developed by T.S.. All co-authors discussed the data. T.S., B.G and W.D. wrote the manuscript.

%\bibliography{InSe}

\end{document}